\documentclass[aps,twocolumn,showpacs,preprintnumbers,floats]{revtex4} 

\usepackage{amsfonts}

\usepackage{graphicx}
\usepackage{dcolumn}
\usepackage{bm}
\usepackage{epsfig}
\newcommand{\vsig}{\mbox{\boldmath$\sigma$\unboldmath}}

\begin{document}

\title{Strong decays of the newly observed $D(2550)$,
$D(2600)$, $D(2750)$ and $D(2760)$}
\author{
Xian-Hui Zhong \footnote {E-mail: zhongxh@ihep.ac.cn}}

\affiliation{ Department of Physics, Hunan Normal University, and
Key Laboratory of Low-Dimensional Quantum Structures and Quantum
Control of Ministry of Education, Changsha 410081, P.R. China }


\begin{abstract}

The strong decay properties of the newly observed $D(2550)$,
$D(2600)$, $D(2750)$ and $D(2760)$ are studied in a constituent
quark model. It is predicted that the $D(2760)$ and $D(2750)$ seem
to be two overlapping resonances. The $D(2760)$ could be identified
as the $1^3D_3$ with $J^P=3^-$, while the $D(2750)$ is most likely
to be the high-mass mixed state $|1{D_2}'\rangle_H$ ($J^P=2^-$) via
the $1^1D_2$-$1^3D_2$ mixing. The $D(2600)$ favors the low-mass
mixed state $|(SD)_1\rangle_L$ $(J^P=1^-)$ via the $1^3D_1$-$2^3S_1$
mixing. The $D(2550)$ as the $2^1S_0$ assignment bears controversies
for its too broad width given in experiments.

\end{abstract}


\maketitle

\section{Introduction}\label{intr}

Recently, four new charmed mesons, $D(2550)$, $D(2600)$, $D(2750)$
and $D(2760)$, were observed by BaBar
Collaboration~\cite{Benitez:2010}. The $D(2600)^0$ and $D(2760)^0$
with neutral charge were first found in the $D^+\pi^-$ channel. Then
their isospin partners $D(2600)^+$ and $D(2760)^+$ were observed in
$D^0\pi^+$ as well. Further analysis of the $D^{*+}\pi^-$ invariant
mass spectrum confirmed the $D(2600)^0$. Furthermore, two additional
new charmed mesons, $D(2550)^0$ and $D(2750)^0$, were found in the
$D^{*+}\pi^-$ channel. The measured branching ratio fractions are
\begin{eqnarray}
\frac{D(2600)^0\rightarrow D^+\pi^-}{D(2600)^0\rightarrow
D^{*+}\pi^-}
=0.32\pm 0.02_{\mathrm{stat}}\pm 0.09_{\mathrm{syst}},\\
\frac{D(2760)^0\rightarrow D^+\pi^-}{D(2750)^0\rightarrow
D^{*+}\pi^-} =0.42\pm 0.05_{\mathrm{stat}}\pm 0.11_{\mathrm{syst}}.
\end{eqnarray}
The other observed results are summarized in Tab.~\ref{wfS}. To
determine the spin-parity $J^P$ of these newly observed charmed
mesons, the BaBar Collaboration also analyzed their helicity
distributions.

These newly observed charmed mesons make great progress in the
establishment of the charmed meson spectroscopy. From the PDG
book~\cite{Amsler:2008zzb}, it is seen that only six low-lying
states, $D$, $D^*$, $D_0(2400)$, $D_1(2430)$, $D_1(2420)$ and
$D_2(2460)$, have been established. The higher excitations, $2S$ and
$1D$ waves, are still absent. Thus, the find of $D(2550)$,
$D(2600)$, $D(2750)$ and $D(2760)$ provides us a good opportunity to
establish the missing $2S$ and $1D$ states.

The $D(2550)^0$ may be identified as the radial excitation of the
$D^0$ (i.e. $2^1S_0$)~\cite{Benitez:2010}, for its quark model
predicted mass $\sim 2.58$ GeV~\cite{Godfrey:1985xj, Di
Pierro:2001uu,Ebert:2009ua}, helicity distribution ($\propto \cos^2
\theta_H$)~\cite{Benitez:2010} and dominated decay mode $D^*\pi$
consist with the observations.

The $D(2600)$ is observed in both $D\pi$ and $D^*\pi$ channels,
thus, its possible $J^P$ are $1^-$ and $3^-$ in the $2S$ and $1D$
states. The BaBar analysis of helicity distribution ($\propto \sin^2
\theta_H$)~\cite{Benitez:2010} also indicates that the $D(2600)$ may
be $1^-$ or $3^-$ assignments~\cite{Godfrey:1985xj}. The typical
quark model predicted mass of $1^3D_3$ is $\sim 2.83$
GeV~\cite{Godfrey:1985xj, Di Pierro:2001uu}, which is much larger
than that of $D(2600)$. Thus, the $D(2600)$ as the $J^P=3^-$
assignment should be excluded.

The $D(2750)$ and $D(2760)$ may be good candidates of $D$ wave
states for their masses are close to those of $D$ waves predicted in
various quark models~\cite{Godfrey:1985xj, Di
Pierro:2001uu,Ebert:2009ua}. Since the $D(2750)^0$ is observed in
$D^{*+}\pi^-$ channel, its possible $J^P$ are $1^-$, $2^-$ and
$3^-$. The helicity distribution of the $J^P=1^-$ and $3^-$
assignments is a simple $\sin^2\theta_H$
distribution~\cite{Godfrey:1985xj}, which is inconsistent with the
BaBar observation that the $D(2750)^0$ does not show a simple
helicity distribution~\cite{Benitez:2010}. Although the mass of
$D(2760)$ is very close to that of $D(2750)$, they may be two
different states for their mass and width values differ by
2.6$\sigma$ and 1.5$\sigma$, respectively~\cite{Benitez:2010}. The
observation of $D(2760)$ in $D\pi$ channel indicates it may be a
candidate of $1^3D_1$ or $1^3D_3$.

To distinguish the different candidates for these newly observed
charmed mesons, in this work, we study their strong decay properties
in a constituent quark model, which has been developed and
successfully used to deal with the strong decays of heavy-light
mesons and charmed baryons~\cite{Zhong:2009sk, Zhong:2008kd,
Zhong:2007gp}. Very recently, the strong decays of the $D(2550)$,
$D(2600)$ and $D(2760)$ were studied by Liu \emph{et al.} in a
$^3P_0$ model~\cite{Sun:2010pg}. For the $D(2550)$ and $D(2600)$,
the main $^3P_0$ model predictions are compatible with our quark
model predictions. In~\cite{Sun:2010pg}, two candidates are
suggested for the $D(2760)$. They are the mixed state via
$2^3S_1$-$1^3D_1$ mixing and $1^3D_3$, respectively. In our
predictions, only the $1^3D_3$ is the favored assignment to
$D(2760)$.

The paper is organized as follows. In the subsequent section, a
brief review of the model is given. The numerical results are
presented and discussed in Sec.~\ref{cwr}. Finally, a summary is
given in Sec.\ \ref{suma}.

\begin{table}[ht]
\caption{Summary of the experimental results.} \label{wfS}
\begin{tabular}{|c|c|c|c|c|c|c }\hline\hline
resonance & mass & width &  decay channel   \\
\hline
$D(2550)^0$ & $2539.4\pm4.5\pm 6.8 $ & $130\pm 12\pm13$ &$D^{*+}\pi^-$       \\
\hline
$D(2600)^0$ & $2608.7\pm 2.4\pm 2.5$ & $93\pm 6\pm13$ &$D^+\pi^-$,$D^{*+}\pi^-$       \\
\hline
$D(2760)^0$ & $2763.3\pm 2.3\pm 2.3$ & $60.9\pm 5.1\pm 3.6$ &$D^+\pi^-$       \\
\hline
$D(2750)^0$ & $2752.4\pm 1.7\pm 2.7$ & $71\pm 6\pm11$      &$D^{*+}\pi^-$     \\
\hline
$D(2600)^+$ & $2621.3\pm 3.7\pm 4.2$ & $93$ &$D^0\pi^+$       \\
\hline
$D(2760)^+$ & $2769.7\pm 3.8\pm 1.5$ & $60.9    $ &$D^0\pi^+$      \\
\hline
\end{tabular}
\end{table}

\section{The model}\label{cr}

In the chiral quark model~\cite{Manohar:1983md}, the low energy
quark-pseudoscalar-meson interactions in the SU(3) flavor basis are
described by the effective
Lagrangian~\cite{Li:1994cy,Li:1997gda,qk3}
\begin{equation}\label{coup}
{\cal L}_{Pqq}=\sum_j
\frac{1}{f_m}\bar{\psi}_j\gamma^{j}_{\mu}\gamma^{j}_{5}\psi_j\partial^{\mu}\phi_m,
\end{equation}
where $\psi_j$ represents the $j$-th quark field in the hadron,
$\phi_m$ is the pseudoscalar meson field, and $f_m$ is the
pseudoscalar meson decay constant.

The effective Lagrangian for quark-vector-meson interactions in the
SU(3) flavor basis is~\cite{zhao:1998fn,Zhao:2000tb,Zhao:2001jw}
\begin{equation}
{\cal
L}_{Vqq}=\sum_j\bar{\psi}_j(a\gamma^{j}_{\mu}+\frac{ib}{2m_j}\sigma_{\mu\nu}q^\nu)V^\mu\psi_j
\ ,
\end{equation}
where $V^\mu$ represents the vector meson field with four-vector
moment $q$. Parameters $a$ and $b$ denote the vector and tensor
coupling strength, respectively.

To match the non-relativistic harmonic oscillator wave function of
the heavy-light meson $\psi^n_{lm}=R_{nl}Y_{lm}$ adopted in the
calculation of the strong decay amplitudes, we should provide the
quark-pseudoscalar and quark-vector-meson coupling operators in a
non-relativistic form. Considering light meson emission in a
heavy-light meson strong decays, the effective
quark-pseudoscalar-meson coupling operator in the center-of-mass
system of the initial meson is~\cite{Zhong:2008kd, Zhong:2009sk,
Li:1994cy,Li:1997gda,qk3}
\begin{eqnarray}\label{ccpk}
H_{m}=\sum_j\left[A \vsig_j \cdot \textbf{q}
+\frac{\omega_m}{2\mu_q}\vsig_j\cdot \textbf{p}_j\right]I_j
\varphi_m ,
\end{eqnarray}
where $A\equiv -(1+\frac{\omega_m}{E_f+M_f})$. In a case when a
light vector meson is emitted, the transition operators for
producing a transversely or longitudinally polarized vector meson
are as follows~\cite{zhao:1998fn,Zhao:2000tb,Zhao:2001jw}:
\begin{eqnarray}\label{vc}
H_m^T=\sum_j \left\{i\frac{b'}{2m_q}\vsig_j\cdot
(\mathbf{q}\times\mathbf{\epsilon})+\frac{a}{2\mu_q}\mathbf{p}_j\cdot
\mathbf{\epsilon}\right\}I_j\varphi_m
\end{eqnarray}
and
\begin{eqnarray}
H_m^L=\sum_j \frac{a M_v}{|\mathbf{q}|}I_j\varphi_m \ .
\end{eqnarray}
In the above three equations,  $\textbf{q}$ and $\omega_m$ are the
three-vector momentum and energy of the final-state light meson,
respectively. $\textbf{p}_j$ is the internal momentum operator of
the $j$-th quark in the heavy-light meson rest frame. $\vsig_j$ is
the spin operator on the $j$-th quark of the heavy-light system, and
$\mu_q$ is a reduced mass given by $1/\mu_q=1/m_j+1/m'_j$ with $m_j$
and $m'_j$ for the masses of the $j$-th quark in the initial and
final mesons, respectively. Here, the $j$-th quark is referred to as
the active quark involved at the quark-meson coupling vertex. $M_v$
is the mass of the emitted vector meson. The plane wave part of the
emitted light meson is $\varphi_m=e^{-i\textbf{q}\cdot
\textbf{r}_j}$, and $I_j$ is the flavor operator defined for the
transitions in the SU(3) flavor space
\cite{Li:1997gda,qk3,zhao:1998fn,Zhao:2000tb,Zhao:2001jw,Zhong:2009sk,
Zhong:2008kd, Zhong:2007gp}. The parameter $b'$ in Eq.(\ref{vc}) is
defined as $b'\equiv b-a$.

For a light pseudoscalar meson emission in a heavy-light meson
strong decays, the partial decay width can be calculated with
\begin{equation}\label{dww}
\Gamma=\left(\frac{\delta}{f_m}\right)^2\frac{(E_f+M_f)|\textbf{q}|}{4\pi
M_i(2J_i+1)} \sum_{J_{iz},J_{fz}}|\mathcal{M}_{J_{iz},J_{fz}}|^2 ,
\end{equation}
where $\mathcal{M}_{J_{iz},J_{fz}}$ is the transition amplitude,
$J_{iz}$ and $J_{fz}$ stand for the third components of the total
angular momenta of the initial and final heavy-light mesons,
respectively. $\delta$ as a global parameter accounts for the
strength of the quark-meson couplings. In the heavy-light meson
transitions, the flavor symmetry does not hold any more. Treating
the light pseudoscalar meson as a chiral field while treating the
heavy-light mesons as constituent quark system is an approximation.
This will bring uncertainties to coupling vertices and form factors.
The parameter $\delta$ is introduced to take into account such an
effect. It has been determined in our previous study of the strong
decays of the charmed baryons and heavy-light mesons
\cite{Zhong:2007gp,Zhong:2008kd}. Here, we fix its value the same as
that in Refs.~\cite{Zhong:2008kd,Zhong:2007gp}, i.e. $\delta=0.557$.

In the calculation, the standard quark model parameters are adopted.
Namely, we set $m_u=m_d=330$ MeV, $m_s=450$ MeV, and $m_c=1700$ MeV
for the constituent quark masses. The harmonic oscillator parameter
$\beta$ in the wave function $\psi^n_{lm}=R_{nl}Y_{lm}$ is taken as
$\beta=0.40$ GeV. The decay constants for $\pi$, $K$ and $\eta$
mesons are taken as $f_{\pi}=132$ MeV, $f_K=f_{\eta}=160$ MeV,
respectively. For the quark-vector-meson coupling strength which
still suffers relatively large uncertainties, we adopt the values
extracted from vector meson photoproduction, i.e. $a\simeq -3$ and
$b'\simeq
5$~\cite{zhao:1998fn,Zhao:2000tb,Zhao:2001jw,Hleiqawi:2007ad,Nanova:2008kr}.
The masses of the mesons used in the calculations are adopted from
the PDG~\cite{Amsler:2008zzb}. With these parameters, the strong
decay properties of the well known heavy-light mesons and charmed
baryons have been described reasonably~\cite{Zhong:2009sk,
Zhong:2008kd, Zhong:2007gp}.

Our approach is similar to Pierro and Eichten's model ~\cite{Di
Pierro:2001uu} in the calculation of the strong decay. Both of the
models adopt the chiral quark-pseudoscalar-meson interactions in the
quark model framework. On the other hand, there are obvious
differences between these two models. Our model is a
non-relativistic quark model, where the non-relativistic harmonic
oscillator wave function of the heavy-light meson is adopted, with
which the decay amplitudes can be presented analytically. Pierro and
Eichten's model is a relativistic quark model, in which the total
wave function is obtained by solving the relativistic Dirac equation
for the heavy-light system.

\section{RESULTS AND DISCUSSIONS}\label{cwr}

\subsection{$D(2550)$} \label{cy}

\begin{table}[ht]
\caption{The partial decay widths and total width (MeV) for the
$D(2550)$ as the $2^1S_0$ candidate, where the mass of $D_0(2400)$
is set with 2338 MeV~\cite{Benitez:2010}.} \label{2550}
\begin{tabular}{|c|c|c|c|c|c|c|c|c|c|c|c|c|c| }\hline\hline
         & $D^*\pi$&$D_0(2400)\pi$& total & $\Gamma(D_0(2400)\pi)/\Gamma(D^*\pi)$  \\
\hline
$2^1S_0$ & 7.2    & 14.9 & 22.1 &  2.1  \\
\hline
\end{tabular}
\end{table}

The $D(2550)^0$ is observed in $D^{*+}\pi^-$ channel with a broad
width $\Gamma\simeq 130$ MeV~\cite{Benitez:2010}. The decay modes,
the BaBar analysis of angle distributions, and the predicted mass of
various theoretical models~\cite{Godfrey:1985xj,Di
Pierro:2001uu,Ebert:2009ua} indicate that it should be classified as
the $2^1S_0$. If $D(2550)$ is considered as the $2^1S_0$ assignment,
it has two decay modes $D^*\pi$ and $D_0(2400)\pi$. The calculated
partial decay widths and total width are listed in Tab.~\ref{2550},
which shows that the predicted width $\Gamma\simeq 22$ MeV is too
narrow to compare with the data. The $^3P_0$ model~\cite{Sun:2010pg}
and relativistic quark model~\cite{Di Pierro:2001uu} calculations
also predicted that the $2^1S_0$ is a narrow width state. The width
of $D(2550)$ may be overestimated if it is the $2^1S_0$ assignment
indeed. To confirm $D(2550)$, further experimental study is needed.

\begin{widetext}
\begin{center}
\begin{table}[ht]
\caption{The partial decay widths and total width (MeV) for
$D(2600)$ as the $2^3S_1$ and $1^3D_1$ candidates, respectively. }
\label{260}
\begin{tabular}
{|c|c|c|c|c|c|c|c|c|c|c|c|c|c| }\hline\hline
         & $D\pi$ & $D_sK$ & $D\eta$ &$D^*\pi$&$D^*\eta$ & $D^*_sK$&$D_1(2430)\pi$&$D_1(2420)\pi$
         &$D_2(2460)\pi$& total   \\
\hline
$2^3S_1$ & 1.9    & 2.4 &2.7    & 9.9 & 1.3&0.02&23.3&0.01& 0.002& 41.5     \\
\hline
$1^3D_1$ & 119.9    & 17.9 &23.1    & 39.0 & 1.8 &0.03&7.9 &43.6& 0.00& 253.2     \\
\hline
\end{tabular}
\end{table}
\end{center}
\end{widetext}

\subsection{$D(2600)$} \label{cz}

\begin{center}
\begin{figure}[ht]
\centering \epsfxsize=8 cm \epsfbox{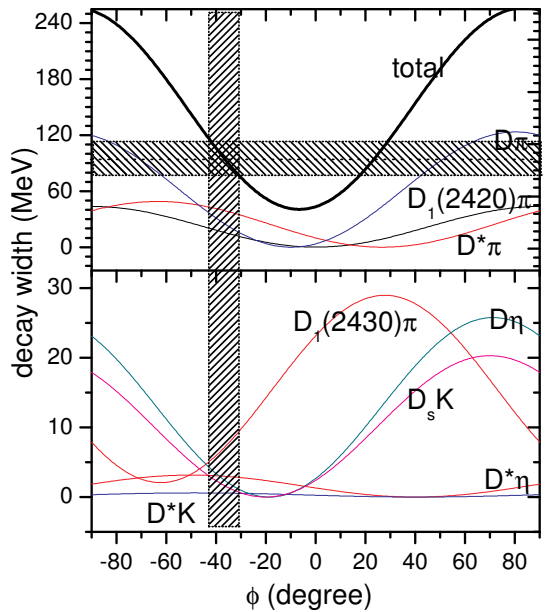} \caption{ (Color
online) The partial decay widths and total decay width of
$|(SD)_1\rangle_L$ with a mass of 2609 MeV as a function of mixing
angle $\phi$. For the tiny contributions of the $D_2(2460)\pi$ and
$D^*_sK$, they are not shown in the figure.}\label{fig-1}
\end{figure}
\end{center}

The $D(2600)$ is observed in both $D\pi$ and $D^*\pi$
channels~\cite{Benitez:2010}. Our analysis in Sec.~\ref{intr}
suggests its quantum number should be $J^P=1^-$. There are two
states, $2^3S_1$ and $1^3D_1$, with $J^P=1^-$ in the $S$ and $D$
waves. The quark model predicted masses of $2^3S_1$ and $1^3D_1$ are
around 2.6 GeV and $2.76$ GeV, respectively~\cite{Godfrey:1985xj,Di
Pierro:2001uu}. The $2^3S_1$-$1^3D_1$ mixing is also possible, for
their comparable masses.

First, we consider $D(2600)$ as the $2^3S_1$ assignment. The decay
modes and corresponding partial decay widths are listed in
Tab.~\ref{260}. The strong decays of this state are dominated by
$D_1(2430)\pi$ and $D^*\pi$. The total decay width and the partial
decay width ratio between $D\pi$ and $D^*\pi$ channels are
\begin{eqnarray}
\Gamma\simeq 42\ \mathrm{MeV},\ \
\frac{\Gamma(D\pi)}{\Gamma(D^*\pi)}\simeq 0.2.
\end{eqnarray}
It shows that the predicted width $\Gamma\simeq 42$ MeV is too
narrow to compare with the data although the ratio
$\Gamma(D\pi)/\Gamma(D^*\pi)$ is compatible with that of
measurement. Thus, with the pure $2^3S_1$ we can not well explain
observations of $D(2600)$. Our conclusion is consistent with that of
$^3P_0$ model~\cite{Sun:2010pg}. Furthermore, the relativistic quark
model calculations also indicate that the $2^3S_1$ is a narrow width
state (with the determined value $g^8_A=0.53\sim 0.82$, the
predicted decay width is $\Gamma\simeq (23\sim 57)$ MeV)~\cite{Di
Pierro:2001uu}. The strong decay properties of $2^3S_1$ in $D$
mesons were studied by Colangelo \emph{et al.} as well with the
heavy quark effective theory~\cite{Colangelo:2007ds}. In their
framework, when the $D(2600)$ is considered as the $2^3S_1$
assignment its decay width, $\Gamma\simeq (128\pm 61)$ MeV, is
compatible with that of measurement, while the predicted ratio,
$\Gamma(D\pi)/\Gamma(D^*\pi)\simeq 0.82$, is obviously larger than
the measured value $\Gamma(D\pi)/\Gamma(D^*\pi)=0.32\pm 0.02\pm
0.09$.

Since $D(2600)$ can not be well explained  with the pure $2^3S_1$
assignment, we consider the possibility of $D(2600)$ as the
$1^3D_1$, the predicted partial widths and total width are shown in
Tab.~\ref{260} as well. It is seen that the predicted width
$\Gamma\simeq 250$ MeV is about a factor 3 larger than the data,
while the predicted ratio $\Gamma(D\pi)/\Gamma(D^*\pi)\simeq  3.1$
is also inconsistent with the data. Thus, the possibility of
$D(2600)$ as the pure $1^3D_1$ is excluded as well.

Finally, we consider the possibility of $D(2600)$ as a mixed state
via the $2^3S_1$-$1^3D_1$ mixing. For which the physical states can
be expressed as
\begin{eqnarray}\label{eq5}
|(SD)_1\rangle_L=+\cos (\phi) |2^3S_1\rangle+\sin(\phi)|1^3D_1\rangle, \\
|(SD)'_1\rangle_H=-\sin (\phi)
|2^3S_1\rangle+\cos(\phi)|1^3D_1\rangle \label{eq6} ,
\end{eqnarray}
where the physical partner in the mixing is included. Assuming that
the low-mass state $|(SD)_1\rangle_L$ corresponds to $D(2600)$, we
plot the decay width of $|(SD)_1\rangle_L$ as a function of the
mixing angle $\phi$ in Fig.~\ref{fig-1}. It is shown that when we
take the mixing angle $\phi\simeq-(36\pm 6)^\circ$, the measured
decay width
\begin{eqnarray}
\Gamma\simeq (93\pm 6\pm 13)\ \mathrm{MeV}
\end{eqnarray}
can be well explained. The predicted partial width ratio is
\begin{eqnarray}
\frac{\Gamma(D\pi)}{\Gamma(D^*\pi)}\simeq 0.63\pm 0.21,
\end{eqnarray}
which is compatible with the measurement ratio
$\Gamma(D\pi)/\Gamma(D^*\pi)=0.32\pm 0.02\pm 0.09$ within its
uncertainties. Thus, the $D(2600)$ may be identified as the mixed
state $|(SD)_1\rangle_L$. Its main strong decay channels are
$D^*\pi$, $D\pi$, $D_1(2420)\pi$ and $D_1(2430)\pi$.

Recently, $D(2600)$ as an admixture of $2^3S_1$ and $1^3D_1$ has
also been suggested by Liu \emph{et al.}~\cite{Sun:2010pg}. They
adopted different mixing scheme from ours. In our mixing scheme
their predicted mixing angle, $-86^\circ\leq \phi \leq -51^\circ$,
is roughly comparable with our prediction $\phi\simeq-(36\pm
6)^\circ$. However, we have noted that the ratio
$\Gamma(D\pi)/\Gamma(D^*\pi)\simeq 2.13\sim 2.86$ predicted by Liu
\emph{et al.}~\cite{Sun:2010pg} is too large to compare with the
observation $\Gamma(D\pi)/\Gamma(D^*\pi)=0.32\pm 0.02\pm 0.09$.

It should be mentioned that in our previous
work~\cite{Zhong:2009sk}, we have discussed the $2^3S_1$-$1^3D_1$
mixing in the study of the $D_{sJ}$ mesons. We predicted that the
$D_s(2710)$ is most likely to be the low-mass state
$|(SD)_1\rangle_L$ with a mixing angle $\phi\simeq-(54\pm 7)^\circ$,
similar prediction also were obtained
in~\cite{Close:2006gr,Li:2009qu}. This mixing angle is close to that
of $D(2600)$. If both $D(2600)$ and $D_s(2710)$ correspond to the
the mixed state $|(SD)_1\rangle_L$ indeed, the $2^3S_1$-$1^3D_1$
mixing might be a common character in the heavy-light mesons. The
future search for $|(SD)_1\rangle_L$ in $B$ and $B_s$ spectroscopies
will clarify this assumption. Finally, we should point out that
there still exist controversies in $D_s(2710)$ about the extent of
the mixing. The $D_s(2710)$ is also interpreted as the first radial
excitation of $D_s^*$ (i.e. $2^3S_1$)~\cite{Colangelo:2007ds}, which
just corresponds to the limit of zero mixing of $|(SD)_1\rangle_L$.
A combined study of $D(2600)$ and $D_s(2710)$ may be helpful to
clarify these controversies.

\begin{center}
\begin{figure}[ht]
\centering \epsfxsize=8 cm \epsfbox{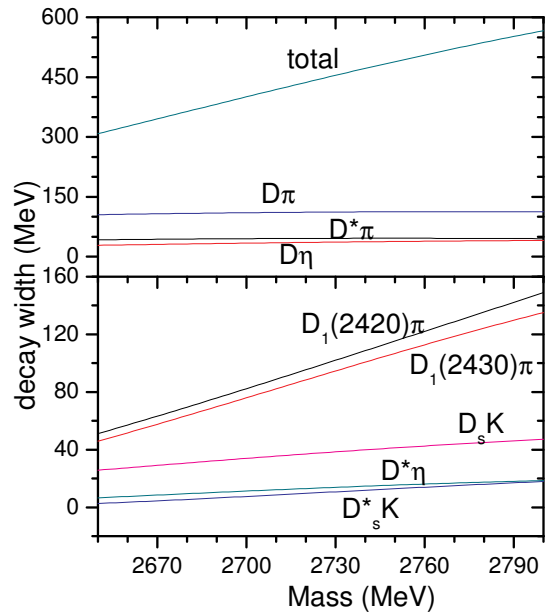} \caption{ (Color
online) The partial decay widths and total width of
$|(SD)'_1\rangle_H$ as a function of mass with the mixing angle
$\phi=-36^\circ$. The tiny contributions of the $D\rho$, $D\omega$
and $D_2(2460)\pi$ are not shown in the figure. }\label{fig-2}
\end{figure}
\end{center}

Following this mixing scheme, one can examine the high-mass partner
$|(SD)'_1\rangle_H$. Supposing that the mass of $|(SD)'_1\rangle_H$
in the range of $(2.65\sim 2.80)$ GeV, in Fig.~\ref{fig-2} we plot
the decay width as a function of the mass with the mixing angle
$\phi=-36^\circ$ fixed by $D(2600)$. It is shown that the
$|(SD)'_1\rangle_H$ should be a broad state with a width of
$\Gamma=(300\sim 550)$ MeV. Its decay modes are dominated by $D\pi$
and $D^*\pi$ , with the increasing mass, the $D_1(2420)\pi$ and
$D_1(2430)\pi$ decay channels become dominant as well.

\subsection{$D(2760)$} \label{czz}

The $D(2760)$ is a good candidate of $D$ waves~\cite{Benitez:2010},
in which the $J^P=2^-$ states [i.e. $^1D_2(2^-)$ and $^3D_2(2^-)$]
are excluded for the observation of the $D\pi$ decay mode. Thus,
only the $1^3D_1(1^-)$ and $1^3D_3(3^-)$ are possible candidates for
$D(2760)$. Assuming the $D(2760)$ as a candidate of $1^3D_1(1^-)$ or
$1^3D_3(3^-)$, it can decay into $D\pi$, $D_sK$, $D\eta$, $D^*\pi$,
$D^*\eta$, $D^*_sK$, $D_1(2430)\pi$, $D_1(2420)\pi$, $D_2(2460)\pi$,
$D\omega$ and $D\rho$. We calculate these partial decay widths and
list the results in Tab.~\ref{276}.

\begin{widetext}
\begin{center}
\begin{table}[ht]
\caption{The decay partial decay widths and total width (MeV) for
$D(2760)$ as the $1^3D_3$ and $1^3D_1$ candidates, respectively.}
\label{276}
\begin{tabular}
{|c|c|c|c|c|c|c|c|c|c|c|c|c|c| }\hline\hline
         & $D\pi$ & $D_sK$ & $D\eta$ &$D^*\pi$&$D^*\eta$ & $D^*_sK$&$D_1(2430)\pi$&$D_1(2420)\pi$
         &$D_2(2460)\pi$&$D\omega$&$D\rho$& total   \\
\hline
$1^3D_3$ & 32.5    & 2.1 &2.6    & 20.6 & 0.7&0.3&5.2&1.7& 1.7&0.1&0.4 & 67.9     \\
\hline
$1^3D_1$ & 156.8    & 45.8 &43.2    & 64.9 & 12.9&10.3&29.4&187.1& 2.7&0.05&0.2 & 553.3     \\
\hline
\end{tabular}
\end{table}
\end{center}
\end{widetext}

As the assignment of $1^3D_1(1^-)$, from the table it is seen that
the strong decays of $D(2760)$ are dominated by $D\pi$ and
$D_1(2420)\pi$. The dominant roles of the $D\pi$ and $D_1(2420)\pi$
decay modes in the strong decays of $1^3D_1(1^-)$ were also
predicted in~\cite{Di Pierro:2001uu, Close:2005se}. It is found that
the total decay width, $\Gamma\simeq 550$ MeV, is too broad to
compare with the data. Thus, $D(2760)$ as the $1^3D_1(1^-)$
assignment should be excluded.

As the assignment of $1^3D_3(3^-)$, the $D(2760)$ has two dominant
decay channels $D\pi$ and $D^*\pi$, which is compatible with the
predictions in~\cite{Di Pierro:2001uu,Close:2005se}. The other decay
modes, such as $D_1(2430)\pi$, $D_sK$ and $D\eta$ have sizeable
contributions. The decay width and partial decay width ratio are
\begin{eqnarray}
\Gamma\simeq 68\ \mathrm{MeV},\ \
\frac{\Gamma(D\pi)}{\Gamma(D^*\pi)}\simeq1.58.
\end{eqnarray}
Our predicted ratio is compatible with the ratio
$\Gamma(D\pi)/\Gamma(D^*\pi)\simeq 1.36$ predicted in~\cite{Di
Pierro:2001uu}, while our predicted width $\Gamma\simeq 68$ MeV is
in agreement with the data $\Gamma\simeq 60.9$ MeV. Furthermore, the
typical quark model predicted mass of $1^3D_3(3^-)$ is $\sim 2.8$
GeV~\cite{Di Pierro:2001uu, Godfrey:1985xj}, which is close to the
mass of $D(2760)$. Thus, the $D(2760)$ is most likely to be the
$1^3D_3(3^-)$ assignment.

Finally, it should be mentioned that in Ref.~\cite{Sun:2010pg} two
possible assignments to $D(2760)$ are suggested, which are
$1^3D_3(3^-)$ and the high-mass partner $|(SD)'_1\rangle_H$ via the
$2^3S_1$-$1^3D_1$ mixing, respectively. Our calculations exclude
$D(2760)$ as the $|(SD)'_1\rangle_H$ assignment. It is shown in
fig.~\ref{fig-2} that as the assignment of $|(SD)'_1\rangle_H$, the
$D(2760)$ should be a broad resonance with a width of $\Gamma\simeq
500$ MeV. The $D\pi$, $D_1(2420)\pi$, $D_1(2430)\pi$, $D^*\pi$ and
$D\eta$ are the main decay modes. For the too broad decay width to
compare with the data, the $D(2760)$ as a mixed state of
$2^3S_1$-$1^3D_1$ is excluded. The differences in the predicted
width of $|(SD)'_1\rangle_H$ between our model and that in
Ref.~\cite{Sun:2010pg} mainly come from the different predictions of
the strong decay properties of $1^3D_1$. In our model, the decays of
the $|(SD)'_1\rangle_H$ are dominated by both the $D\pi$ and
$D_1(2420)\pi$ channels. We find that the main contributor to the
partial widths of $D\pi$ and $D_1(2420)\pi$ is the $1^3D_1$, whose
decay modes are dominated by $D\pi$ and $D_1(2420)\pi$. However, in
Ref.~\cite{Sun:2010pg} the strong decays of $1^3D_1$ are predicted
to be dominated by $D_1(2430)\pi$. It should be pointed out that
with the $^{3}P_0$ model, Close and Swanson predicted that the
dominant decay modes of $1^3D_1$ are $D_1(2420)\pi$ and
$D\pi$~\cite{Close:2005se}. In fact, it is easy to distinguish the
two different assignments to the $D(2760)$ in experiments by
measuring the ratio $\Gamma(D^*\pi)/\Gamma(D\pi)$, for its very
different value in the two cases.

\begin{center}
\begin{figure}[ht]
\centering \epsfxsize=8 cm \epsfbox{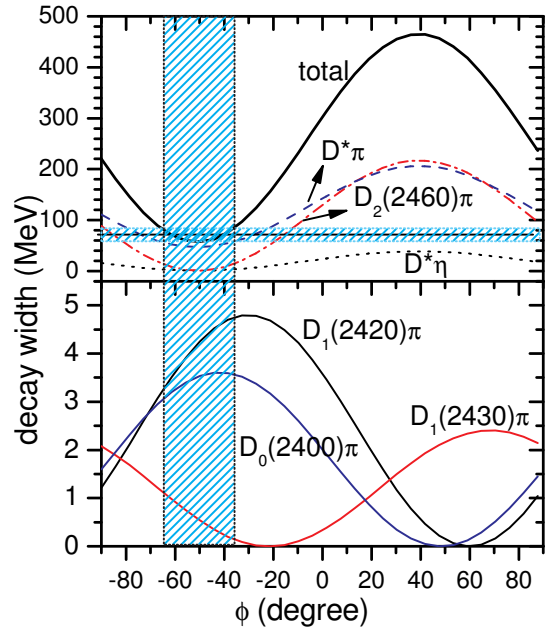} \caption{ (Color
online) The partial decay widths and total width of
$|1{D_2}'\rangle_H$ with a mass of 2750 MeV as a function of mixing
angle $\phi$. The tiny contributions of the $D\rho$ and $D\omega$,
are not shown in the figure. }\label{fig-4}
\end{figure}
\end{center}

\subsection{$D(2750)$} \label{crz}

The $D(2750)^0$ is observed in $D^{*+}\pi^-$. Although its mass is
very close to that of $D(2760)$, they might be two different
resonances due to the following three reasons: (i) If they are the
same charmed meson state, according to our analysis in the
Sec.~\ref{czz} they should be the $1^3D_3$ assignment. However, the
simple helicity distribution of $1^3D_3$, $\propto \sin^2
\theta_H$~\cite{Godfrey:1985xj}, is inconsistent with the
observation that the $D(2750)^0$ does not show a simple helicity
distribution~\cite{Benitez:2010}; (ii) Furthermore, the predicted
ratio $\Gamma(D\pi)/\Gamma(D^*\pi)\simeq 1.58$ is inconsistent with
the measured value $\Gamma(D\pi)/\Gamma(D^*\pi)\simeq 0.42$ if they
are the same state; (iii) Their measured mass and width values
differ by 2.6$\sigma$ and 1.5$\sigma$,
respectively~\cite{Benitez:2010}. The recent study of the strong
decays of $D(2750)$ and $D(2760)$ with the heavy quark effective
theory agrees with our conclusion~\cite{wangzg}.

Thus, the $D(2750)$ is most likely to be the $J^P=2^-$ assignments.
There are three cases, $1^1D_2$, $1^3D_2$ and their admixtures of
$1^1D_2$-$1^3D_2$, should be considered. First, we consider the
$D(2750)^0$ as a mixed state of $1^1D_2$-$1^3D_2$ by the following
mixing scheme:
\begin{eqnarray}\label{eq2}
|1D_2\rangle_L=+\cos (\phi) |1^1D_2\rangle+\sin(\phi)|1^3D_2\rangle, \\
|1{D_2}'\rangle_H=-\sin
(\phi)|1^1D_2\rangle+\cos(\phi)|1^3D_2\rangle\label{eq3}  ,
\end{eqnarray}
where the subscripts $L$ and $H$ denote the low-mass and high-mass
state due to the mixing. Usually, the $|1{D_2}'\rangle_H$ has a
narrow width~\cite{Godfrey:1986wj,Close:2005se,Swanson}. We thus
consider the $D(2750)$  as the $|1{D_2}'\rangle_H$. In
Fig.~\ref{fig-4} the decay properties of $|1{D_2}'\rangle_H$ as a
function of the mixing angle $\phi$ are plotted. We see that when we
take the mixing angle $\phi\simeq -(50\pm 15)^\circ$, the predicted
decay width is in the range of BaBar observation $\Gamma=(71\pm 6\pm
11)$ MeV. The decay modes are dominated by the $D^*\pi$, which can
explain why the $D(2750)^0$ is first observed in $D^{*+}\pi^-$
channel. It is also interestedly found that the mixing angle is
consistent with that ($\phi=50.7^\circ$) obtained in the heavy quark
effective
theory~\cite{Ebert:2009ua,Close:2005se,Godfrey:1986wj,Swanson}.
Considering the $D(2760)$ as the $1^3D_3$, we predicted the ratio
\begin{eqnarray}
\frac{D(2760)\rightarrow D\pi}{D(2750)\rightarrow
D^*\pi}\simeq0.37\sim 0.57,
\end{eqnarray}
which is in good agreement with the observed value as well. As a
whole the $D(2750)$ is favorably interpreted as the mixed state
$|1{D_2}'\rangle_H$ with a mixing angle $\phi\simeq -(50\pm
15)^\circ$. The $D(2750)$ might be observed in $D_1(2420)\pi$,
$D_0(2400)\pi$, $D^*\eta$ and $D_1(2430)\pi$ channels for their
sizeable partial widths.

The $D(2750)$ can not be interpreted as either a pure $1^1D_2$ state
or a pure $1^3D_2$ state for their too broad widths to compare with
the data. It is shown in Fig.~\ref{fig-4}, the decay widths of the
$1^1D_2$ and $1^3D_2$ are $\Gamma\simeq 220$ MeV (taking
$\phi=90^\circ$) and $\Gamma\simeq 330$ MeV (taking $\phi=0^\circ$),
respectively.

\begin{center}
\begin{figure}[ht]
\centering \epsfxsize=8.5 cm \epsfbox{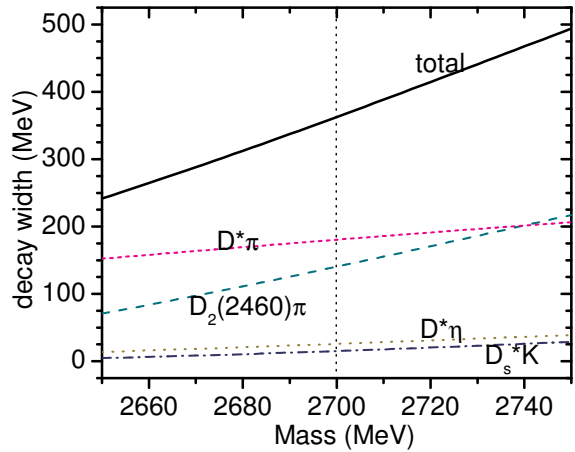} \caption{ (Color
online) The partial decay widths and total width of $|1D_2\rangle_L$
as a function of mass with the mixing angle $\phi=-50^\circ$. For
the tiny contributions of the $D\rho$, $D\omega$, $D_0(2400)\pi$,
$D_1(2420)\pi$ and $D_1(2430)\pi$, they are not shown in the figure.
}\label{fig-5}
\end{figure}
\end{center}

Since the $D(2750)$ can be interpreted as the mixed state
$|1{D_2}'\rangle_H$, its low-mass partner $|1D_2\rangle_L$ may be
observed in experiments as well. It is predicted that the mass of
low-mass partner $|1D_2\rangle_L$ is about 50 MeV lighter than that
of $|1{D_2}'\rangle_H$~\cite{Ebert:2009ua}. Thus, the mass of
$|1D_2\rangle_L$ is likely to be $\sim 2.7$ GeV. To know about the
decay properties of $|1D_2\rangle_L$, in Fig.~\ref{fig-5} we plot
its decay width as a function of mass in the range of $(2.65\sim
2.75)$ GeV with a mixing angle $\phi=-50^\circ$ fixed by $D(2750)$.
From the figure we see that the $|1D_2\rangle_L$ should be a broad
state with a width of $\Gamma\simeq (250\sim 500)$ MeV. Its strong
decays are dominated by $D^*\pi$ and $D_2(2460)\pi$. Furthermore,
the $D^*\eta$ and $D_s^*K$ also have sizeable contributions to the
strong decays of $|1D_2\rangle_L$. The $|1D_2\rangle_L$ may be too
broad to be observed in experiments.

\subsection{Sensitivity to $\beta$} \label{cz}

\begin{center}
\begin{figure}[ht]
\centering \epsfxsize=9 cm \epsfbox{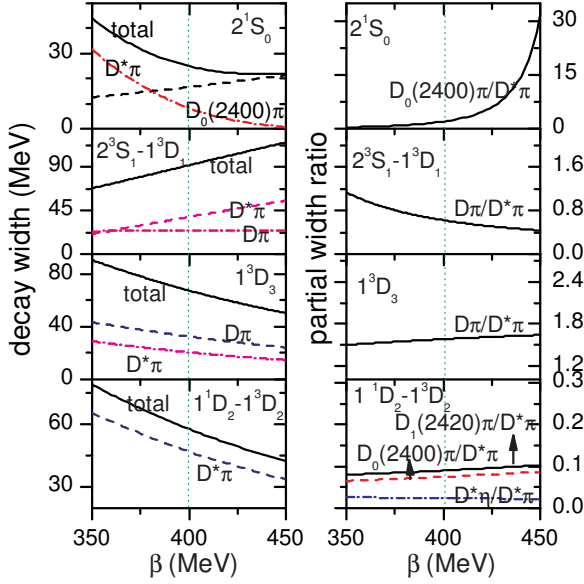} \caption{ (Color
online) The partial decay widths, total widths and partial decay
width ratios of different configuration assignments as a function of
$\beta$, which have been labeled in the figure, where we only plot
the dominant decay channels of these assignments. The
$2^3S_1-1^3D_1$ and $2^3S_1-1^3D_1$ stand for the mixed states
$|(SD)_1\rangle_L$ and $|1{D_2}'\rangle_H$, respectively. The mixing
angle of $|(SD)_1\rangle_L$ is fixed with $\phi=-36^\circ$, while
the mixing angle of $|1{D_2}'\rangle_H$ is set with
$\phi=-50^\circ$. The masses of $2^1S_0$, $|(SD)_1\rangle_L$,
$|1{D_2}'\rangle_H$ and $1^3D_3$ are set with $2539$ MeV, 2609 MeV,
2750 MeV and 2760 MeV, respectively. }\label{fig-3}
\end{figure}
\end{center}

The harmonic oscillator parameter $\beta$ is the most important
parameter in the quark model. It controls the size effect or
coupling form factor from the convolution of the heavy-light meson
wave functions. The uncertainties of $\beta$ may affect our
conclusions. The typical quark model value of $\beta$ is $\sim 0.4$
GeV. To examine the sensitivity of the calculation to $\beta$, we
plot the decay widths, partial decay widths and partial decay width
ratios of $2^1S_0$, $1^3D_3$, mixed state $|(SD)_1\rangle_L$ of
$2^3S_1$-$1^3D_1$ and mixed state $|1{D_2}'\rangle_H$ of
$1^3D_2$-$1^1D_2$ as a function $\beta$ in Fig.~\ref{fig-3}.

It shows that the decay widths of these excited charmed mesons
exhibit some sensitivities to the parameter $\beta$. The
uncertainties of the width of $1^3D_3$ mainly come from the $D\pi$
and $D^*\pi$ channels, while for the $2^1S_0$, the
$|(SD)_1\rangle_L$ and the $|1{D_2}'\rangle_H$, the uncertainties of
their decay widths mainly come from the $D^*\pi$ channel. Within the
range of $\beta=(400\pm 50)$ MeV, about a 30\% uncertainty of the
decay widths would be expected, which consists with our previous
analysis~\cite{Zhong:2009sk}. This is a typical order of accuracy
for the constituent quark model, and can be regarded as reasonable.

From the figure, we see that the ratios
$\Gamma(D_0(2400)\pi)/\Gamma(D\pi)$ of $2^1S_0$  and
$\Gamma(D\pi)/\Gamma(D^*\pi)$ of the mixed state of
$|(SD)_1\rangle_L$ are sensitive to $\beta$. In contrast, the ratios
of the $D$ waves, $1^3D_3$ and $|1{D_2}'\rangle_H$, are insensitive
to $\beta$.

In brief, although the harmonic oscillator parameter $\beta$ can
bring some uncertainties to the final results, within the range of
$\beta=(400\pm 50)$ MeV, our major conclusions will still hold.

Finally, it should be mentioned that the relatively large
uncertainties of the quark-vector-meson couplings, $a$ and $b'$,
might affect our conclusions as well. Fortunately, they only affect
the decay channels of a light-vector meson emission, such as $D\rho$
and $D\omega$ channels. From the Tab.IV, we see that although the
$D\rho$ and $D\omega$ are allowed for $D(2750, 2760)$, their partial
decay widths predicted in our model are so tiny that we can neglect
their contributions. In fact when we use large values for the
quark-vector-meson couplings, $a$ and $b'$, the partial widths of
$D\rho$ and $D\omega$ are still small. Thus, here we do not consider
the effects of their uncertainties on the results.

\section{summary}\label{suma}

\begin{table}[ht]
\caption{$D$ and $D_s$ meson spectroscopies. The $1 P_1(1^+)$ and $1
P'_1(1^+)$ stand for the mixed states via the $1^1P_1$-$1^3P_1$
mixing defined in Refs.~\cite{Close:2005se}. The
$|(SD)_1\rangle_L(1^-)$ and $|(SD)'_1\rangle_H(1^-)$ are the mixed
states via the $2^3S_1$-$1^3D_1$ mixing defined in Eqs. (\ref{eq5})
and (\ref{eq6}), respectively, while the $|1D_2\rangle_L(2^-)$ and
$|1D'_2\rangle_H(2^-)$ are the mixed states via the
$1^3D_2$-$1^1D_2$ mixing defined in Eqs. (\ref{eq2}) and
(\ref{eq3}), respectively.} \label{dds}
\begin{tabular}{|c|c|c|c|c|c|c|c|c|c|c|c|c|c }\hline\hline
 $n ^{2S+1}L_J(J^P)$   & $D_J$ state & $D_{sJ}$ state   \\
\hline
$1^1S_0(0^-)$ & D(1865)    & $D_s(1968)$     \\
\hline
$1^3S_1(1^-)$ & $D^*(2007)$    & $D_s(2112)$      \\
\hline
$1^3P_0(0^+)$ & $D_0(2400)$   & $D_{s0}(2317)$      \\
\hline
$1 P_1(1^+)$  &  $D_1(2430)$    & $D_{s1}(2460)$     \\
\hline
$1 P'_1(1^+)$ & $D_1(2420)$   & $D_{s1}(2536)$      \\
\hline
$1^3P_2(2^+)$ & $D_2(2460)$   & $D_{s2}(2573)$      \\
\hline
$2^1S_0(0^-)$ & $D(2550)$?   & ?        \\
\hline
$|(SD)_1\rangle_L(1^-)$ & $D(\mathbf{2600})$    & $D_s(\mathbf{2710})$     \\
\hline
$|(SD)'_1\rangle_H(1^-)$ &     ?    &    ?       \\
\hline
$|1D_2\rangle_L(2^-)$ & ?    & ?      \\
\hline
$|1D'_2\rangle_H(2^-)$ & $D(\mathbf{2750})$    & $D_{sJ_2}(\mathbf{2860})$      \\
\hline
$1^3D_3(3^-)$ & $D(\mathbf{2760})$    & $D_{sJ_1}(\mathbf{2860})$        \\
\hline
\end{tabular}
\end{table}

In this work we have studied the strong decay properties of the
newly observed $D(2550)$, $D(2600)$, $D(2750)$ and $D(2760)$ by
BaBar Collaboration in a constituent chiral quark model. These newly
observed charmed mesons provide us a chance to establish a more
completed $D$ meson spectroscopy, which has been shown in
Tab.~\ref{dds}. For comparison, the $D_s$ meson spectroscopy is also
included.

We have found that  $D(2550)^0$ as the $2^1S_0$ is still
questionable. The predicted narrow width of $2^1S_0$ is inconsistent
with the observation, although its decay modes, helicity
distributions and theoretical predicted mass satisfy this
classification. Given the poor statistics of $D(2550)^0$, its decay
width may be overestimated by experimentalists. We expect them to
observe it in both $D^*\pi$ and $D_0(2400)\pi$ channels.

The $D(2600)$ can be identified as the low mass mixed state
$|(SD)_1\rangle_L(1^-)$ via the $2^3S_1$-$1^3D_1$ mixing. This mixed
state is also predicted in the $D_s$ meson spectroscopy, which
corresponds to the $D_s(2710)$~\cite{Zhong:2009sk}. In our mixing
scheme the high mass partner $|(SD)_1\rangle_H(1^-)$ may be too
broad to be observed in $D$ meson spectroscopy, while it might be
found in $D_s$ spectroscopy~\cite{Zhong:2009sk}. To understand the
nature of $D(2600)$ further, we suggest to observe it in
$D_1(2420)\pi$, $D_1(2430)\pi$, $D\eta$ and $D_sK$ channels.

The $D(2760)$ is most likely to be the $1^3D_3(3^-)$. Its decays are
governed by $D\pi$ and $D^*\pi$, which can naturally explain why
$D(2760)$ is first observed in $D\pi$ channel. The predicted ratio
is $\Gamma(D\pi)/\Gamma(D^*\pi)\simeq 1.58$. The $D(2760)$ as high
mass partner of $D(2600)$ via the $2^3S_1$-$1^3D_1$ mixing was also
suggested in~\cite{Sun:2010pg}, where the predicted partial decay
width of $D\pi$ is tiny. Further experimental measurement of the
ratios $\Gamma(D^*\pi)/\Gamma(D\pi)$, $\Gamma(D_sK)/\Gamma(D\pi)$
and $\Gamma(D\eta)/\Gamma(D\pi)$ should be able to disentangle its
properties.

The $D(2750)$ and $D(2760)$ might be two different charmed meson
states. The $D(2750)$ is favorably interpreted as the high mass
mixed state $|1{D_2}'\rangle_H$ ($2^-$) via the $1^1D_2$-$1^3D_2$
mixing. Its low-mass partner $|1{D_2}\rangle_L$ may be too broad to
be observed in experiments. To confirm the $D(2750)$, the decay
channels $D_1(2420)\pi$, $D_0(2400)\pi$, $D^*\eta$ and
$D_1(2430)\pi$ are suggested to be observed in future experiments.

Finally, we should mention that in our previous
work~\cite{Zhong:2009sk}, we predicted that $D_s(2860)$ might
correspond to two largely overlapping resonances, one resonance is
likely to be the $1^3D_3$ [denoted by $D_{sJ_1}(2860)$] and the
other resonance seems to be the mixed state between $1^3D_2$ and
$1^1D_2$ [denoted by $D_{sJ_2}(2860)$]. Combining the study of the
$D(2750)$ and $D(2760)$ in present work, we easily conclude that
both $D(2760)$ and $D_{sJ_1}(2860)$ are most likely to be the
$1^3D_3$, while both $D(2750)$ and $D_{sJ_2}(2860)$ might be
classified as the mixed state $|1{D_2}'\rangle_H$ with almost the
same mixing angle. To test our predictions and clarify the
controversial situation of $D_s(2860)$~\cite{Colangelo:2006rq,
Chen:2009zt,vanBeveren:2009jq,Li:2009qu}, we suggest to analyze the
helicity distribution of $D_{s}(2860)\rightarrow D^*K$ in
experiments. If the helicity distribution is in proportion to
$(1+h\cos^2\theta_H)$, the $D_s(2860)$ should be two largely
overlapping resonances.


\section*{  Acknowledgements }

The author would like to thank Professor Xiang Liu for drawing his
attention to the BaBar Collaboration's new data. This work is
supported by the National Natural Science Foundation of China (Grant
Nos. 10775145 and 11075051).

\end{document}